\documentclass[prd,twocolumn,showpacs,floatfix,amsmath,amssymb,floatfix,nofootinbib]{revtex4}
\usepackage{graphicx,dcolumn,booktabs,bm}
\usepackage{longtable,lscape}
\usepackage{txfonts}
\usepackage{overpic}
\usepackage{amssymb}
\usepackage{indentfirst}
\usepackage{feynmf}   
\usepackage{slashed}  
\usepackage{cases}
\usepackage{color,ulem}

\begin{document}


\title{First estimate of producing the charmed baryon $\Lambda_c(2880)$ at $\overline{\mbox{P}}$ANDA}
\author{Qing-Yong Lin$^{1,2,3}$}\email{qylin@impcas.ac.cn}
\author{Xiang Liu$^{1,4}$\footnote{Corresponding author}}\email{xiangliu@lzu.edu.cn}
\author{Hu-Shan Xu$^{1,2}$}

\affiliation{
$^1$Research Center for Hadron and CSR Physics, Lanzhou University and Institute of Modern Physics of CAS, Lanzhou 730000, China\\
$^2$Institute of Modern Physics, Chinese Academy of Sciences, Lanzhou 730000, China\\
$^3$University of Chinese Academy of Sciences, Beijing, 100049, China\\
$^4$School of Physical Science and Technology, Lanzhou University, Lanzhou 730000, China}

\date{\today}

\begin{abstract}
In the present work we explore the production potential of $\Lambda_c(2880)^+$ at PANDA. With the $J^P=\frac{5}{2}^+$ assignment to $\Lambda_c(2880)^+$, we calculate the differential and total cross sections of $p\bar{p}\to \Lambda_c^-\Lambda_c(2880)^+$. We also perform the Dalitz plot analysis and give the $pD^0$ invariant mass spectrum distribution of the process $p\bar{p}\to \Lambda_c^-pD^0$, where the signal and background contributions are considered. Our numerical results indicate that the production of $\Lambda_c(2880)^+$ may reach up to about 20 $\mu$b. About $10^7$ events from the reconstruction of $pD^0$ can be accumulated per day, if taking the designed luminosity ($2\times10^{32}~\mbox{cm}^{-2}\mbox{s}^{-1}$) of PANDA.
\end{abstract}

\maketitle

\section{Introduction}\label{sec:intro}

The higher orbital excitation of the $\Lambda_c$ baryon family, $\Lambda_c(2880)$ was first announced by the CLEO Collaboration through analyzing the $M(\Lambda_c^+\pi^+\pi^-)-M(\Lambda_c^+)$ mass difference plot \cite{Artuso:2000xy}. In 2007, Belle carried out the study of  $\Lambda_c(2880)$, where  $\Lambda_c(2880)$ decay into $\Sigma_c(2455)^{0,++}\pi^{+,-}$ was observed. What is more important is that the measurement of its spin-parity assignment was given, i.e., its $J^P$ favors $5/2^+$ \cite{Abe:2006rz}. In Ref. \cite{Aubert:2006sp}, $\Lambda_c(2880)\to D^0p$ was reported by the BABAR Collaboration. The resonance parameters of $\Lambda_c(2880)$ include \cite{Abe:2006rz,Aubert:2006sp,Beringer:1900zz}
\begin{eqnarray*}
\mathrm{Belle}:\, M&=&2881.2\pm0.2\pm0.4\, \mathrm{MeV},\\\Gamma&=&5.8\pm0.7\pm1.1\, \mathrm{MeV},\\
\mathrm{BABAR}:\, M&=&2881.9\pm0.1\pm0.5\, \mathrm{MeV},\\ \Gamma&=&5.8\pm1.5\pm1.1 \,\mathrm{MeV}.
\end{eqnarray*}
The above experimental phenomena show that the experimental information of $\Lambda_c(2880)$ is quite abundant among all observed charmed baryons \cite{Beringer:1900zz}.

After the observation of $\Lambda_c(2880)$, different theoretical groups have performed theoretical studies of $\Lambda_c(2880)$. Most of the theoretical studies of $\Lambda_c(2880)$ mainly focus on the decay behavior of $\Lambda_c(2880)$  as we are going to introduce.
By the quark pair creation model, the strong decay behaviors of charmed baryons are investigated systematically \cite{Chen:2007xf}. The results indicate that $\Lambda_c(2880)$ favors $\check{\Lambda}_{c3}^2(5/2^+)$ (the notation of the charmed baryon can be found in Fig. 3 of Ref. \cite{Chen:2007xf}) since the corresponding total decay width and the ratio $\Gamma(\Sigma_c(2520)\pi)/\Gamma(\Sigma_c(2455)\pi)$ are consistent with the experimental data of $\Lambda_c(2880)$ given by Belle \cite{Abe:2006rz}.
In Ref. \cite{Cheng:2006dk}, Cheng and Chua calculated the strong decays of $\Lambda_c(2880)$ by the heavy hadron chiral perturbation theory, where $\Lambda_c(2880)$ can be a mixture of $L=2$ charmed baryons $\Lambda_{c2}(5/2^+)$ and $\tilde{\Lambda}_{c3}^{\prime\prime}(5/2^+)$ \cite{Cheng:2006dk}.
Later, Zhong and Zhao also studied the charmed baryon strong decays via a chiral quark model, which contain the
$\Lambda_c(2880)$ two-body strong decay \cite{Zhong:2007gp}.

Although studying the decay behavior of $\Lambda_c(2880)$ is helpful to reveal the inner structure of $\Lambda_c(2880)$, exploring the production of $\Lambda_c(2880)$ is also an important and intriguing research topic. Until now, all experimental observations of $\Lambda_c(2880)$ have been from the $B$ meson weak decays \cite{Artuso:2000xy,Abe:2006rz,Aubert:2006sp}. Thus, it is natural to ask whether $\Lambda_c(2880)$ can be produced by other processes. To answer this question, in this work we will carry out the study of
the $\Lambda_c(2880)$ production.
We notice that $\Lambda_c(2880)$ can decay into $D^0p$ \cite{Aubert:2006sp}, which shows that there exists the strong coupling between $\Lambda_c(2880)$  and  $D^0p$. In addition, searching for the charmed baryon is one of the physical aims at PANDA \cite{Lutz:2009ff}.  Considering the above reasons, we study the discovery potential of $\Lambda_c(2880)$ at PANDA, which can provide valuable information to future experimental exploration of
$\Lambda_c(2880)$ at PANDA.

This work is organized as follows. After the introduction, we present the selected process of $\Lambda_c(2880)$
produced at PANDA and the corresponding calculation detail. In Sec. \ref{sec3}, the Dalitz plot and the $pD^0$ invariant mass spectrum are given, which contains the analysis of the signal and background contributions. Finally, this paper ends with a discussion and conclusion in Sec. \ref{sec4}.

\section{The production of $\Lambda_c(2880)$}\label{sec2}

\begin{figure}[htb]
\begin{center}
\scalebox{0.9}{\includegraphics[width=\columnwidth]{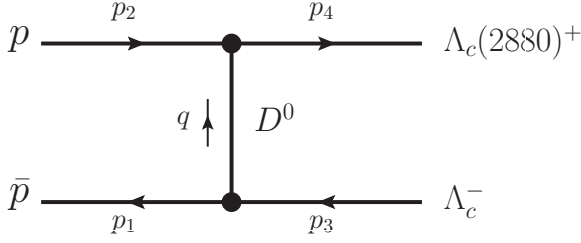}}
\caption{The diagram describing the $p\bar p \to \Lambda_c^-\Lambda_c(2880)^+$ process.
\label{fig:2to2}}
\end{center}
\end{figure}

Since $\Lambda_c(2880)$ can couple with $pD^0$ \cite{Aubert:2006sp}, $\Lambda_c(2880)$ can be produced in the proton and antiproton collision process $p\bar p \to \Lambda_c^-\Lambda_c(2880)^+$ by exchanging a $D^0$ meson, which is shown in Fig. \ref{fig:2to2}. In the present work, we do not consider the contribution from the direct $p\bar{p}$ annihilation, which is suppressed by the Okubo-Zweig-Iizuka rule \cite{Okubo:1963fa,Zweig:1964jf,Iizuka:1966fk,Griffiths:2008zz}.

In the following, we calculate the production probability of $\Lambda_c(2880)$ in the process $p\bar p \to \Lambda_c^-\Lambda_c(2880)^+$ by the effective Lagrangian approach, where the differential and total cross sections are discussed.

\subsection{The Lagrangians and the coupling constants}\label{sec21}

As measured by Belle \cite{Abe:2006rz}, we take the quantum number of $\Lambda_c(2880)^+$ to be $J^{PC}=\frac{5}{2}^+$.
For depicting the coupling of the nucleon with the charmed meson and the charmed baryon, we adopt the effective Lagrangians \cite{Zou:2002yy,Wu:2009md,Ouyang:2009kv}
\begin{eqnarray}
\mathcal{L}_{DN\Lambda_c} &=& -\frac{g_{DN\Lambda_c}}{m_{\Lambda_c}+m_N}\overline{\Lambda}_c\gamma_5\gamma^{\mu} \partial_\mu DN + H.c., \\
\mathcal{L}_{DNR} &=& -\frac{g_{DNR}}{m_R+m_N}\overline{R}^{\mu\nu}\gamma_5\gamma^{\lambda}
(\partial_\lambda\partial_\mu\partial_\nu D)N + H.c.,
\end{eqnarray}
where $N$, $R$, and $D$ are the isodoublet nucleon field, the $\Lambda_c(2880)$ field, and the isodoublet $D$
meson field, respectively, with the definitions $N=(p,n)^T$, $\overline{N}=(\bar{p},\bar{n})$, $D=(D^0,D^+)$,
and $\overline{D}=(\bar{D}^0,D^-)^T$. In the following formulas, we take $g_{\Lambda_c}\equiv g_{DN\Lambda_c}$ and
$g_R\equiv g_{DNR}$ for the sake of convenience.

The propagators for the fermion with $J=1/2,~5/2$ are expressed as \cite{Machleidt:1987hj,Tsushima:1998jz,Huang:2005js,Wu:2009md}
\begin{eqnarray}
G_\mathcal{F}^{n+(1/2)}(p) &=& \tilde{P}^{(n+(1/2))}(p)\frac{i2m_\mathcal{F}}{p^2-m_\mathcal{F}^2+im_\mathcal{F}\Gamma_\mathcal{F}}
\end{eqnarray}
with
\begin{eqnarray}
\tilde{P}^{1/2}(p) &=& \frac{\slashed{p}+m_\mathcal{F}}{2m_\mathcal{F}},\\
\tilde{P}^{5/2}(p) &=& \frac{\slashed{p}+m_\mathcal{F}}{2m_\mathcal{F}}G_{\mu\nu\alpha\beta}(p),\label{eq:proj5}
\end{eqnarray}
\begin{eqnarray}
G_{\mu\nu\alpha\beta}(p) &=& \frac{1}{2}(\tilde{g}_{\mu\alpha}\tilde{g}_{\nu\beta} +
\tilde{g}_{\mu\beta}\tilde{g}_{\nu\alpha}) - \frac{1}{5}\tilde{g}_{\mu\nu}\tilde{g}_{\alpha\beta} \nonumber\\
 && +\frac{1}{10}(\tilde{\gamma}_{\mu}\tilde{\gamma}_{\alpha}\tilde{g}_{\nu\beta} +
\tilde{\gamma}_{\nu}\tilde{\gamma}_{\beta}\tilde{g}_{\mu\alpha} \nonumber\\
 && +\tilde{\gamma}_{\mu}\tilde{\gamma}_{\beta}\tilde{g}_{\nu\alpha} +
\tilde{\gamma}_{\nu}\tilde{\gamma}_{\alpha}\tilde{g}_{\mu\beta}),
\end{eqnarray}
where $\tilde{g}_{\mu\nu}=-g_{\mu\nu}+p_\mu p_\nu/p^2$ and $\tilde{\gamma}_{\mu}=-\gamma_\mu+\slashed{p}p_\mu/p^2$. In addition, $p$ and $m_\mathcal{F}$ are the momentum and the mass of the fermion, respectively.

By an approximate SU(4) flavor symmetry, the coupling constant
$g_{\Lambda_c}$ is equal to $g_{\Lambda NK}=13.2$ \cite{Rijken:1998yy,Stoks:1999bz,Oh:2006hm,Liu:2010um},
which is larger than $g_{\Lambda NK}=6.7\pm2.1$ estimated by the QCD sum rules \cite{Navarra:1998vi,Bracco:1999xe}.
The former one is adopted in this work. Additionally, the coupling constant $g_{R}$ can be obtained by fitting the
measured partial width of the $\Lambda_c(2800)^+(Q) \to D^0(K)p(P)$ decay, where the partial decay width is
\begin{eqnarray}
d\Gamma_i &=& \frac{m_Rm_N}{8\pi^2}|\mathcal{M}|^2\frac{|\vec{K}|}{m_R^2}d\Omega
\end{eqnarray}
with
\begin{eqnarray}
E_K &=& \frac{m_R^2-m_N^2+m_D^2}{2m_R}, \\
|\vec{K}| &=& \frac{\sqrt{(m_R^2-(m_D+m_N)^2)(m_R^2-(m_D-m_N)^2)}}{2m_R}.
\end{eqnarray}
Here, $E_K$ and $\vec{K}$ are the energy and the three-momentum of the daughter $D^0$ meson, respectively.
$m_N$ and $m_D$ are the masses of proton and $D^0$ meson, respectively.
Furthermore, the concrete expression of the corresponding decay width is
\begin{eqnarray}
\Gamma_i &=& \frac{m_R m_N|\vec{K}|}{2\pi m_R^2}\frac{1}{6} \sum|T_{fi}|^2 \nonumber\\
 &=& \frac{g_R^2 m_N|\vec{K}|}{12\pi m_R(m_N+m_R)^2}\sum Tr[u(P)\bar{u}(P)\gamma_5\slashed{K} \nonumber\\
 && \times K_{\mu}K_{\nu}u_R^{\mu\nu}(Q)\bar{u}_R^{\alpha\beta}(Q)\gamma_5\slashed{K}K_{\alpha}K_{\beta}] \nonumber\\
 &=& \frac{g_R^2|\vec{K}|}{24\pi m_R(m_N+m_R)^2} Tr[(\slashed{P}+m_N)\gamma_5\slashed{K}K_{\mu}K_{\nu} \nonumber\\
 && \times \tilde{P}^{5/2}(Q)\gamma_5\slashed{K}K_{\alpha}K_{\beta}],
\end{eqnarray}
where $\tilde{P}^{5/2}(Q)$ is the projection operator for a fermion with $J=5/2$ as
defined in Eq. (\ref{eq:proj5}). Since the measurement of
the branching ratio of  $\Lambda_c(2880)^+ \to D^0p$ is still absent, we can, thus, determine the coupling constant $g_{R}$ by the theoretical result of the branching ratio of
$\Lambda_c(2880)^+ \to D^0p$. In Ref. \cite{Chen:2007xf},  the estimated branching fraction $BR(\Lambda_c(2880)^+ \to D^0p)$ is around 20\%, where the quark  pair creation model is adopted. The corresponding partial decay width is 1.2 MeV. Considering the above situation, in this work we take typical value $\Gamma(\Lambda_c(2880)^+ \to D^0p)=1$ MeV to extract $g_{R}=40.69$ GeV$^{-2}$, which will be applied to the following calculation.

\begin{figure}[htb]
\begin{center}
\scalebox{0.9}{\includegraphics[width=\columnwidth]{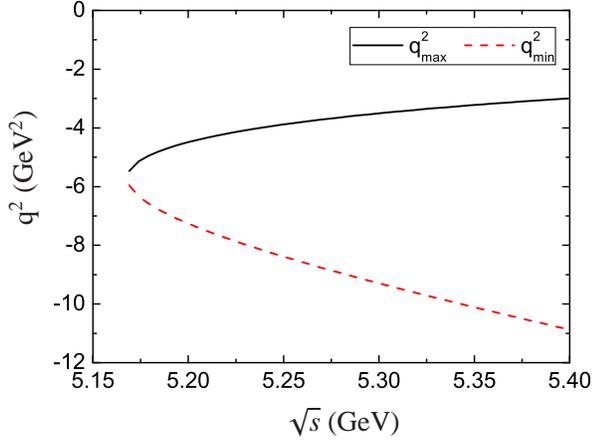}}
\caption{(color online) The kinematically allowed region for the momentum of the transfer momentum in the processes $p\bar p \to \Lambda_c^-\Lambda_c(2880)^+$.
\label{fig:qSuqre}}
\end{center}
\end{figure}

Before carrying out the study of the cross section of $p\bar{p}\to\Lambda_c^-\Lambda_c(2880)^+$, we display the kinematically allowed region of the square of the transfer momentum $q^2$ (see Fig. \ref{fig:qSuqre}), which is the function of $\sqrt{s}$. In Fig. \ref{fig:qSuqre}, the maximum of $q^2$ is negative and less than the mass square of the exchanged $D^0$ meson in the energy range of our interest.

\subsection{The production of $\Lambda_c(2880)$}\label{sec22}

The transition amplitude for the process $p\bar{p} \to \Lambda_c^-\Lambda_c(2880)^+$ shown in Fig. \ref{fig:2to2} is expressed as
\begin{eqnarray}\label{eq:2to2}
i\mathcal{T}_{fi} &=& \frac{g_{\Lambda_c} g_R}{(m_{\Lambda_c}+m_N)(m_N+m_R)}\bar{u}_R(p_4)\mathcal{C}_R(q)u_{p}(p_2) \nonumber\\
&&  \bar{v}_{\bar{p}}(p_1)\mathcal{C}(q)v_{\bar{\Lambda}_c}(p_3)G_D(q^2)\mathcal{F}^2(q^2,m_D^2),
\end{eqnarray}
where $\mathcal{C}(q)=\gamma_5\slashed{q}$ and $\mathcal{C}_R(q)=\gamma_5\slashed{q}q_{\mu}q_{\nu}$
describe the Lorentz structures of the vertices of $D^0$ interacting
with $\bar{p}\bar{\Lambda}_c$ and $p\Lambda_c(2880)^+$, respectively. $G_D(q^2)=i/(q^2-m_D^2)$
is the propagator of the exchanged meson. In addition, the monopole form
factor
\begin{eqnarray}
\mathcal{F}(q^2,m_D^2) &=& \frac{\Lambda^2-m_D^2}{\Lambda^2-q^2}
\end{eqnarray}
is also introduced for the strong interaction vertices, where the phenomenological parameter $\Lambda$ can be parameterized as
\begin{eqnarray}
\Lambda &=& m_D + \alpha \Lambda_{QCD}
\end{eqnarray}
with $\Lambda_{QCD}=$220 MeV, where the dimensionless parameter $\alpha$ is expected to be of order unity \cite{Cheng:2004ru}. Later, we will discuss how to constrain the value of $\alpha$.

The transition amplitude of
$p\bar{p}\to \Lambda_c^-\Lambda_c^+$ can be obtained by replacing $\mathcal{C}_R(q)$ with $-\mathcal{C}(q)$ in Eq. (\ref{eq:2to2}).
The unpolarized cross section is \cite{Beringer:1900zz}
\begin{eqnarray}
\frac{d\sigma}{dt} &=& \frac{m_Nm_Nm_{\Lambda_c}m_R}{16\pi s}
    \frac{1}{|\vec{p}_1|^2}\sum|\mathcal{T}_{fi}|^2,
\end{eqnarray}
where
\begin{eqnarray}
|\mathcal{T}_{fi}|^2 &=& \left(\frac{g_{\Lambda_c} g_R}{(m_{\Lambda_c}+m_N)(m_N+m_R)}\right)^2|G_D(q^2)|^2\mathcal{F}^4(q^2,m_D^2) \nonumber\\
&&  \times Tr\left[P^{5/2}(p_4)\mathcal{C}_R(q)\frac{\slashed{p}_2+m_N}{2m_N}\gamma^0\mathcal{C}_R(q)^\dag\gamma^0\right] \nonumber\\
&&  \times Tr\left[\frac{\slashed{p}_1-m_N}{2m_N}\mathcal{C}\frac{\slashed{p}_3-m_{\Lambda_c}}{2m_{\Lambda_c}}
    \gamma^0\mathcal{C}^\dag\gamma^0 \right].
\end{eqnarray}

\begin{figure}[htb]
\begin{center}
\scalebox{0.9}{\includegraphics[width=\columnwidth]{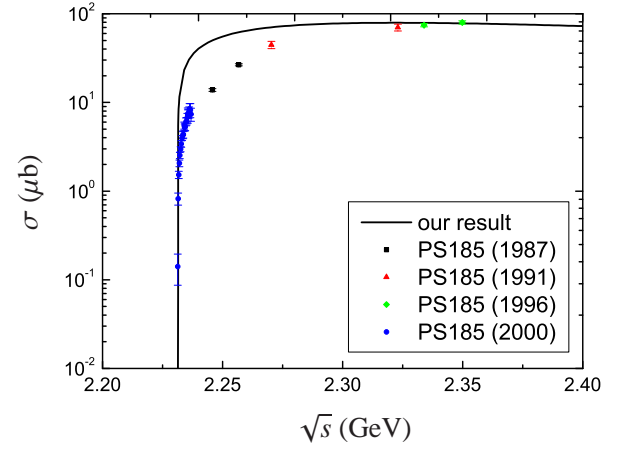}}
\caption{(color online) The obtained total cross section for $p\bar p \to \Lambda^-\Lambda^+$. The data are from the PS185 experiment \cite{Barnes:1987aw,Barnes:1990bs,Barnes:1996cf,Barnes:2000be}.
\label{fig:TCS2toLL}}
\end{center}
\end{figure}

Before giving the numerical results of the production of $\Lambda_c(2880)^+$, we should constrain the value of $\alpha$.

\begin{figure}[htb]
\begin{center}
\scalebox{0.9}{\includegraphics[width=\columnwidth]{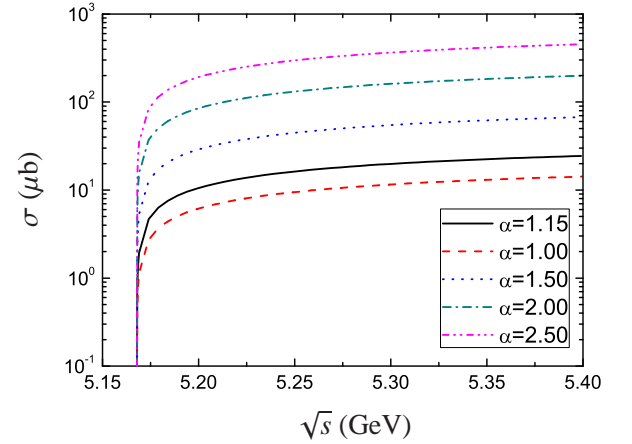}}
\caption{(color online) The obtained total cross section for $p\bar p \to \Lambda_c^-\Lambda_c(2880)^+$ with different $\alpha$, where the result of $\alpha=1.15$ is plotted by the solid line.
\label{fig:TCS2toLR}}
\end{center}
\end{figure}

\begin{figure}[htb]
\begin{center}
\scalebox{0.9}{\includegraphics[width=\columnwidth]{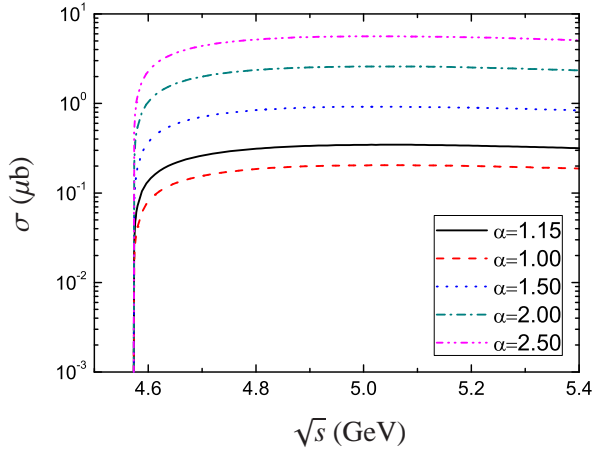}}
\caption{(color online) The obtained total cross section for $p\bar p \to \Lambda_c^-\Lambda_c^+$ with different $\alpha$, where the result of $\alpha=1.15$ is plotted by the solid line.
\label{fig:TCS2toLcLc}}
\end{center}
\end{figure}

We noticed that the reaction $p\bar{p}\to \Lambda\bar{\Lambda}$ has been measured by the PS185 experiment at the Low Energy Antiproton Ring (LEAR), and the data for the corresponding cross section have been reported in Refs. \cite{Barnes:1987aw,Barnes:1990bs,Barnes:1996cf,Barnes:2000be}, which is shown in Fig. \ref{fig:TCS2toLL}. If considering the approximate SU(4) flavor symmetry, $p\bar{p}\to \Lambda_c\bar{\Lambda}_c$ is similar to
$p\bar{p}\to \Lambda\bar{\Lambda}$, which makes us constrain the $\alpha$ value by the experimental data of $p\bar{p}\to \Lambda\bar{\Lambda}$. By extending the formulas of $p\bar{p}\to \Lambda_c\bar{\Lambda}_c$ to the reaction $p\bar{p}\to \Lambda\bar{\Lambda}$, one can easily obtain the corresponding cross section. In Fig. \ref{fig:TCS2toLL}, we show the calculated total cross section of $p\bar{p}\to \Lambda\bar{\Lambda}$ and the comparison with the experimental data. It is obvious that our numerical result can describe the experimental data when $\alpha=1.15$. This $\alpha$ will be applied to obtain the cross sections of $p\bar p \to \Lambda_c^-\Lambda_c(2880)^+$ and $p\bar p \to \Lambda_c^-\Lambda_c^+$ just shown in Figs. \ref{fig:TCS2toLR} and \ref{fig:TCS2toLcLc}, respectively. With $\alpha=1.15$, the cross section of $p\bar{p}\to \Lambda_c^-\Lambda_c(2880)^+$ can reach up to about 20 $\mu$b, while that of $p\bar{p}\to \Lambda_c^-\Lambda_c^+$ is about 0.2 $\mu$b.

In Fig. \ref{fig:TCS2toLR}, the variation of the total cross section of $p\bar{p}\to \Lambda_c^-\Lambda_c(2880)^+$ to the different values of the parameter $\alpha$ is also plotted, where $\alpha$ is taken as $1.0-2.5$ with the step of 0.5. Our results of the $\Lambda_c(2880)^+$ production indicate that the cross section of $p\bar p \to \Lambda_c^-\Lambda_c(2880)^+$ strongly depends on the adopted values of $\alpha$. For example, the cross section of the $\Lambda_c(2880)^+$ production with $\alpha=1.0$ is smaller than that with $\alpha=2.5$. Thus, we need to use the $p\bar{p}\to \Lambda\bar{\Lambda}$ experimental data to constrain the $\alpha$ value. In the following background analysis, we take $\alpha=1.15$ to present the numerical results.

The variation of the total cross section of $p\bar p \to \Lambda_c^+\Lambda_c^-$ to $\alpha$ is also given in Fig. \ref{fig:TCS2toLcLc}. It is obvious that the cross section for this process is rather smaller than that of $p\bar p \to \Lambda_c^-\Lambda_c(2880)^+$.

\begin{figure}[htb]
\begin{center}
\scalebox{0.9}{\includegraphics[width=\columnwidth]{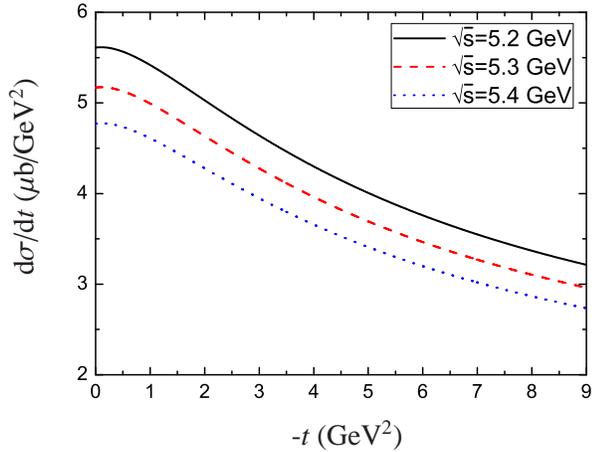}}
\caption{(color online) The differential cross section for $p\bar p \to \Lambda_c^-\Lambda_c(2880)^+$ dependent on $-t$ with the fixed center-of-mass energy $\sqrt{s}=5.2,~5.3,~5.4$ GeV.
\label{fig:DSDT}}
\end{center}
\end{figure}

In addition, we also present the differential cross section of $p\bar p \to \Lambda_c^-\Lambda_c^+(2880)$ with different values of the center-of-mass energy $\sqrt{s}$, which is shown in Fig. \ref{fig:DSDT}.

\section{The background analysis and the Dalitz plot}\label{sec3}
\begin{figure}[htb]
\begin{center}
\scalebox{0.9}{\includegraphics[width=\columnwidth]{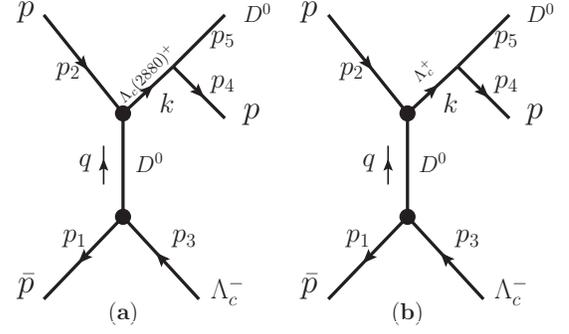}}
\caption{The diagrams for $p\bar{p} \to D^0p\bar{\Lambda}_c$
with the intermediate $\Lambda_c(2880)^+$ (a) and $\Lambda_c^+$ (b) contributions.
\label{fig:2to3}}
\end{center}
\end{figure}

Besides giving the total and differential cross sections of the production of $\Lambda_c(2880)^+$ in the $p\bar{p}$ collision, it is also important to perform the background analysis and the Dalitz plot of the corresponding reaction, which can provide more abundant information of the  $\Lambda_c(2880)^+$ production at PANDA. In the present work, we consider the processes $p\bar{p}\to \Lambda_c^-pD^0$, where $pD^0$ is from the intermediate resonance $\Lambda_c(2880)^+$ or $\Lambda_c^+$ as shown in Fig. \ref{fig:2to3}. The process $p\bar{p}\to\Lambda_c^-\Lambda_c^+\to\Lambda_c^-pD^0$ with the off-shell $\Lambda_c^+$ is as the main background contribution.

The transition amplitudes of $p\bar{p}\to\Lambda_c^-pD^0$ are written as
\begin{eqnarray}
i\mathcal{T}_{fi}^{R} &=& \left(\frac{g_R}{m_N+m_R}\right)^2\frac{g_L}{m_{\Lambda_c}+m_N}\bar{u}_p(p_4)\left[-\mathcal{C}_R(p_5)\right] \nonumber\\
&& \times G_R^{5/2}(k)\mathcal{C}_R(q)u_{p}(p_2) \bar{v}_{\bar{p}}(p_1)\mathcal{C}(q)v_{\bar{\Lambda}_c}(p_3) \nonumber\\
&& \times G_D(q^2)\mathcal{F}^2(q^2,M_D^2),\label{eq:2to3R}
\end{eqnarray}
\begin{eqnarray}
i\mathcal{T}_{fi}^{\Lambda_c} &=& \left(\frac{g_{\Lambda_c}}{m_{\Lambda_c}+m_N}\right)^3\bar{u}_p(p_4)\mathcal{C}(p_5)G_{\Lambda_c}^{1/2}(k) \nonumber\\
&& \times \left[-\mathcal{C}(q)\right]u_{p}(p_2) \bar{v}_{\bar{p}}(p_1)\mathcal{C}(q)v_{\bar{\Lambda}_c}(p_3) \nonumber\\
&& \times G_D(q^2)\mathcal{F}^2(q^2,M_D^2),\label{eq:2to3L}
\end{eqnarray}
which correspond to Figs. \ref{fig:2to3}(a) and \ref{fig:2to3}(b),
where the expressions $\mathcal{C}_R$ and $\mathcal{C}$ are defined in Sec. \ref{sec22}.  The definition of the involved momenta can be found in Fig. \ref{fig:2to3}.

With Eqs. (\ref{eq:2to3R}) and (\ref{eq:2to3L}), one obtains the square of the total invariant transition amplitude
\begin{eqnarray}
|\mathcal{M}|^2 &=& \sum|\mathcal{T}_{fi}^{R}+\mathcal{T}_{fi}^{\Lambda_c}|^2.
\end{eqnarray}
The corresponding total cross section of the process $p\bar{p}\to \Lambda_c^-pD^0$ is
\begin{eqnarray}
 d\sigma &=& \frac{m_N^2}{|p_1\cdot p_2|}\frac{|\mathcal{M}|^2}{4}(2\pi)^4d\Phi_3(p_1+p_2;p_3,p_4,p_5)
\end{eqnarray}
with the definition of the $n$-body phase space \cite{Beringer:1900zz}
$$d\Phi_n(P;k_1,...,k_n)=\delta^4(P-\sum\limits_{i=1}^nk_i)\prod\limits_{i=1}^3\frac{d^3k_i}{(2\pi)^32E_i}.$$

\begin{figure}[htb]
\begin{center}
\scalebox{0.9}{\includegraphics[width=\columnwidth]{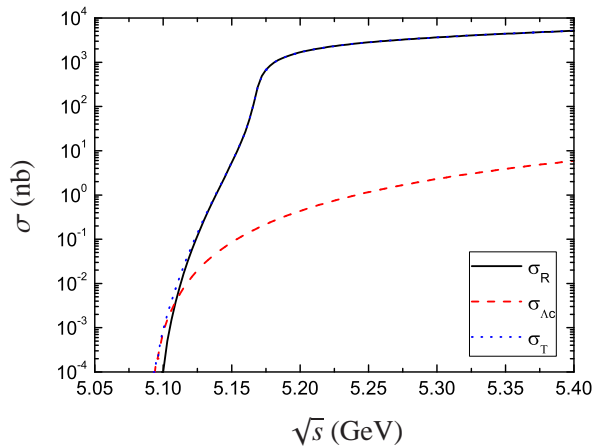}}
\caption{(color online) The obtained total cross section for $p\bar p \to \Lambda_c^-pD^0$. Here, $\sigma_R$ and $\sigma_{\Lambda_c}$ are the results via the exchanged $\Lambda_c(2880)^+$ and $\Lambda_c^+$, respectively, while $\sigma_T$ denotes the total cross section.
\label{fig:TCS2to3}}
\end{center}
\end{figure}

To numerically calculate the total cross section of $p\bar{p}\to \Lambda_c^-pD^0$ including both signal and background contributions, the MATHEMATICA and FOWL codes are utilized. In Fig. \ref{fig:TCS2to3}, the variation of the total cross sections to $\sqrt{s}$ is given, where $\sigma_R$ and $\sigma_{\Lambda_c}$ correspond to the signal and background contributions, respectively. As shown in Fig. \ref{fig:TCS2to3}, the line shape of $\sigma_R$ increases rapidly near the threshold. Since the process can proceed via on-shell intermediate $\Lambda_c(2880)^+$, a steep increase appears at about $\sqrt{s}=5.17$ GeV and  $\sigma_R$ can reach up to about $5$ $\mu$b at the energy range of our interest. $\sigma_{\Lambda_c}$ has a dominant role at $\sqrt{s}<5.11$ GeV. However, $\sigma_R$ becomes important when $\sqrt{s}$ increases, and then $\sigma_R$ is much lager than $\sigma_{\Lambda_c}$ when $\sqrt{s}>5.18$ GeV, which indicates that the signal can be easily distinguished from the background in this energy region.

\begin{figure}[htb]
\begin{center}
\scalebox{0.9}{\includegraphics[width=\columnwidth]{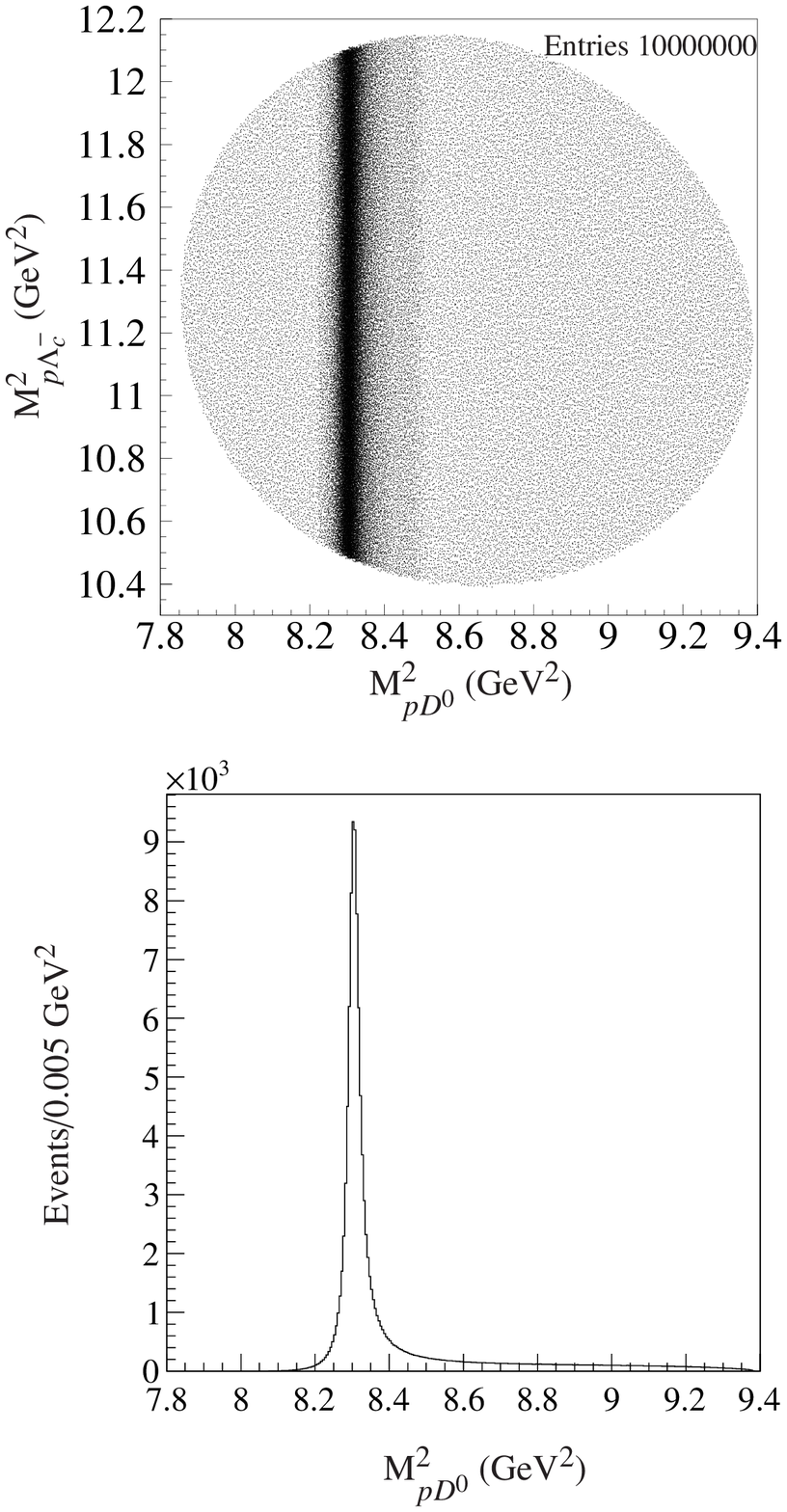}}
\caption{The Dalitz plot (top) and the invariant mass spectrum distribution (bottom) for $p\bar p \to \Lambda_c^-pD^0$ at $\sqrt{s}=5.35$ GeV.
\label{fig:DalitzPlot}}
\end{center}
\end{figure}

After giving the total cross section of $p\bar{p}\to \Lambda_c^-pD^0$, we also carry out the analysis of the Dalitz plot and the $pD^0$ invariant mass spectrum for this process, which are useful for studying the production of $\Lambda_c(2880)^+$ in the proton-antiproton collision $p\bar{p}\to \Lambda_c^-pD^0$. In Fig. \ref{fig:DalitzPlot}, the Dalitz plot and the corresponding $pD^0$ invariant mass spectrum at $\sqrt{s}=5.35$ GeV are given. When $10^7$ events are generated in the Monte Carlo simulation, the signal event can reach up to about $10^4$ events/0.005 GeV$^2$. The $pD^0$ invariant mass spectrum indicates that the signal can be well distinguished from the background. This is due to the fact that the contribution from the signal is far larger than that from the background at the energy range $\sqrt{s}>5.18$ GeV (see Fig. \ref{fig:TCS2to3} for more details).

\section{Discussion and conclusion}\label{sec4}

In this work, we investigate the discovery potential of charmed baryon $\Lambda_c(2880)$ produced at PANDA, which is different from the $\Lambda_c(2880)$ production in $B$ meson decay \cite{Abe:2006rz}. Thus, this study will be helpful to further experimental search for $\Lambda_c(2880)$ at the forthcoming PANDA experiment, where searching for the charmed baryon is one of the most important physical aims of PANDA \cite{Lutz:2009ff}.

The total and differential cross sections of the $\Lambda_c(2880)^+$ production indicate that $p\bar{p}\to \Lambda_c^-\Lambda_c(2880)^+$ is a suitable process to explore the $\Lambda_c(2880)^+$ production at PANDA. Considering the designed luminosity of PANDA ($2\times10^{32}$ $\mbox{cm}^{-2}\mbox{s}^{-1}$), we can estimate that there are about $10^7$ $\Lambda_c(2880)^+$ events accumulated per day by reconstructing the final $pD^0$. In addition, the background analysis and the Dalitz plot are given in this work, where  the process $p\bar{p}\to \Lambda_c^-pD^0$ is calculated by including the signal and background contributions. We find that the contribution from the signal is much larger than that from the background when $\sqrt{s}>5.18$ GeV, which is a suitable energy window to study the $\Lambda_c(2880)^+$ production at PANDA. These results also show that the $\Lambda_c(2880)$ can be easily distinguished from the background.

It should be mentioned that there were some discussions of the initial state interaction (ISI) and final state interaction (FSI) effects on the nucleon-nucleon collisions when the transition occurs near the threshold \cite{Haidenbauer:1991kt,Haidenbauer:1992wp,Kohno:1986mg,Baru:2002rs,Hanhart:2003pg}, where the ISI and FSI effects are thought to be governed by the nonperturbative QCD effects and are rather complicated. The authors of Ref. \cite{Hanhart:1998rn} studied the ISI  effect on meson production in $NN$ collisions, where the ISI leads to a reduction of the total cross section of the order of $|\lambda_L|^2=\eta_L(p)\cos^2(\delta_L(p))+\frac{1}{4}[1-\eta_L(p)]^2$, which relates to the phase shift $\delta_L(p)$ and inelasticity $\eta_L(p)$ of $NN$ scattering. With increasing the energy, the dependence of $|\lambda_L|^2$ on energy becomes smaller.
In addition, by studying the reaction $NN\to NN\eta$, the authors in Ref. \cite{Baru:2002rs} claimed that the FSI effect is not universal and depends on the concrete meson production mechanisms. Usually, the Jost function is utilized to describe the FSI effects \cite{Hinterberger:2005cb}. Haidenbauer et al. also discussed the contribution of FSI to the cross section in the frame of J$\ddot{u}$lich meson-baryon model \cite{Haidenbauer:2009ad}. Their results indicate that the cross sections do not change by more than 10\%-15\% when the ISI effect is included.

However, as the first theoretical estimate of hunting the charmed baryon $\Lambda_c(2880)$ at PANDA, the present work does not seriously consider these effects in studying the production of $\Lambda_c(2880)$ since the ISI and FSI effects are rather complicated, especially for the discussed production process. Further study by including the ISI and FSI effects is an interesting research topic, which should be explored in future work.

In summary, we suggest a future PANDA experiment to perform the search for the charmed baryon $\Lambda_c(2880)^+$. This experimental study cannot only further confirm $\Lambda_c(2880)^+$ by different processes but also provide more abundant information to $\Lambda_c(2880)^+$, which will be valuable to reveal the underlying structure of $\Lambda_c(2880)^+$.

\section{Acknowledgments}

This project is supported
by the National Natural Science Foundation of China under
Grants No. 11222547, No. 11175073, and No. 11035006,
the Ministry of Education of China (FANEDD under Grant
No. 200924, SRFDP under Grant No. 20120211110002,
NCET, the Fundamental Research Funds for the Central
Universities), and the Fok Ying-Tong Education
Foundation (Grant No. 131006).

\vfill

\end{document}